\documentclass{article}
\usepackage{graphicx}
\usepackage{amsfonts}
\usepackage{amsmath}

\begin{document}

\title{Simple models of bouncing ball dynamics and their comparison}
\author{Andrzej Okninski$^{1}$, Bogus\l aw Radziszewski$^{2}$ \\
Physics Division$^{1}$, Department of Mechatronics and \\
Mechanical Engineering$^{2}$, \\
Politechnika Swietokrzyska, Al. 1000-lecia PP 7,\\
25-314 Kielce, Poland}
\maketitle

\begin{abstract}
Nonlinear dynamics of a bouncing ball moving in gravitational field and
colliding with a moving limiter is considered. Several simple models of
table motion are studied and compared. Dependence of displacement of the
table on time, approximating sinusoidal motion and making analytical
computations possible, is assumed as quadratic and cubic functions of time,
respectively.
\end{abstract}

\section{Introduction}

Vibro-impacting systems belong to a very interesting and important class of
nonsmooth and nonlinear dynamical systems \cite%
{diBernardo2008,Luo2006,Awrejcewicz2003,Filippov1988}\ with important
technological applications \cite%
{Stronge2000,Mehta1994,Knudsen1992,Wiercigroch2008}. Dynamics of such
systems can be extremely complicated due to velocity discontinuity arising
upon impacts. A very characteristic feature of such systems is the presence
of nonstandard bifurcations such as border-collisions and grazing impacts
which often lead to complex chaotic motions.

The Poincar\'{e} map, describing evolution from an impact to the next
impact, is a natural tool to study vibro-impacting systems. The main
difficulty with investigating impacting systems is in finding instant of the
next impact what typically involves solving a nonlinear equation. However,
the problem can be simplified in the case of a bouncing ball dynamics
assuming a special motion of the limiter. Bouncing ball models have been
extensively studied, see \cite{Luo2009}\ and references therein. As a
motivation that inspired this work, we mention study of physics and
transport of granular matter \cite{Mehta1994}. A similar model has been also
used to describe the motion of railway bogies \cite{Knudsen1992}.

Recently, we have considered several models of motion of a material point in
a gravitational field colliding with a limiter moving with piecewise
constant velocity \cite{AOBR2008,AOBR2009a,AOBR2009b,AOBR2009c,AOBR2010}. In
the present paper more realistic yet still simple models are considered. The
purpose of this work is to approximate sinusoidal motion as exactly as
possible but still preserving possibility of analytical computations.

The paper is organized as follows. In Section 2 a one dimensional dynamics
of a ball moving in a gravitational field and colliding with a table is
considered and Poincar\'{e} map is constructed. Dependence of displacement
of the table on time is assumed as quadratic and cubic functions of time,
respectively. In Section 3 bifurcation diagrams for such models of table
motion are computed and compared with the case of sinusoidal motion.

\section{Bouncing ball: a simple motion of the table}

We consider a motion of a small ball moving vertically in a gravitational
field and colliding with a moving table, representing unilateral
constraints. The ball is treated as a material point while the limiter's
mass is assumed so large that its motion is not affected at impacts. A
motion of the ball between impacts is described by the Newton's law of
motion:%
\begin{equation}
m\ddot{x}=-mg,  \label{point motion}
\end{equation}%
where $\dot{x}=dx/dt$ and motion of the limiter is:%
\begin{equation}
y=y\left( t\right) ,  \label{limiter motion}
\end{equation}%
with a known function $y$. We shall also assume that $y$ is a continuous
function of time. Impacts are modeled as follows: 
\begin{eqnarray}
x\left( \tau _{i}\right) &=&y\left( \tau _{i}\right) ,  \label{position} \\
\dot{x}\left( \tau _{i}^{+}\right) -\dot{y}\left( \tau _{i}\right)
&=&-R\left( \dot{x}\left( \tau _{i}^{-}\right) -\dot{y}\left( \tau
_{i}\right) \right) ,  \label{velocity}
\end{eqnarray}%
where duration of an impact is neglected with respect to time of motion
between impacts. In Eqs. (\ref{position}), (\ref{velocity}) $\tau _{i}$
stands for time of the $i$-th impact while $\dot{x}_{i}^{-}$, $\dot{x}%
_{i}^{+}$are left-sided and right-sided limits of $\dot{x}_{i}\left(
t\right) $ for $t\rightarrow \tau _{i}$, respectively, and $R$ is the
coefficient of restitution, $0\leq R<1$ \cite{Stronge2000}.

Solving Eq. (\ref{point motion}) and applying impact conditions (\ref%
{position}), (\ref{velocity}) we derive the Poincar\'{e} map \cite{AOBR2007}%
: 
\begin{subequations}
\label{TV}
\begin{eqnarray}
\gamma Y\left( T_{i+1}\right) &=&\gamma Y\left( T_{i}\right) -\Delta
_{i+1}^{2}+\Delta _{i+1}V_{i},  \label{T} \\
V_{i+1} &=&-RV_{i}+2R\Delta _{i+1}+\gamma \left( 1+R\right) \dot{Y}\left(
T_{i+1}\right) ,  \label{V}
\end{eqnarray}%
where $\Delta _{i+1}\equiv T_{i+1}-T_{i}$. The limiter's motion has been
typically assumed in form $Y_{s}(T)=\sin (T)$, cf. \cite{AOBR2009a} and
references therein. This choice leads to serious difficulties in solving the
first of Eqs.(\ref{TV}) for $T_{i+1}$, thus making analytical investigations
of dynamics hardly possible. Accordingly, we have decided to simplify the
limiter's periodic motion to make (\ref{T}) solvable.

In our previous papers we have assumed displacement of the table as the
following periodic function of time: 
\end{subequations}
\begin{equation}
Y\left( T\right) =\left\{ 
\begin{array}{l}
\frac{1}{h}\hat{T},\quad 0\leq \hat{T}<h \\ 
\frac{-1}{1-h}\hat{T}+\frac{1}{1-h},\quad h\leq \hat{T}\leq 1%
\end{array}%
\right.  \label{L}
\end{equation}%
with $\hat{T}=T-\left\lfloor T\right\rfloor $ and $0<h<1$, where $%
\left\lfloor x\right\rfloor $ is the floor function -- the largest integer
less than or equal to $x$. In this work the function $Y(T)$ is assumed as
quadratic $Y_{q}$ and two cubic functions of time, $Y_{c_{1}}$ and $%
Y_{c_{2}} $. More exactly, these functions read:%
\begin{equation}
Y_{q}\left( T\right) =\left\{ 
\begin{array}{l}
-16\hat{T}\left( \hat{T}-\frac{1}{2}\right) ,\quad 0\leq \hat{T}<\frac{1}{2}
\\ 
16\left( \hat{T}-\frac{1}{2}\right) \left( \hat{T}-1\right) ,\quad \frac{1}{2%
}\leq \hat{T}\leq 1%
\end{array}%
\right.  \label{Q}
\end{equation}%
\begin{equation}
Y_{c_{1}}\left( T\right) =12\sqrt{3}\hat{T}\left( \hat{T}-\frac{1}{2}\right)
\left( \hat{T}-1\right) ,\quad 0\leq \hat{T}\leq 1  \label{C1}
\end{equation}%
\begin{equation}
Y_{c_{2}}\left( T\right) =\left\{ 
\begin{array}{cc}
\left( 32\pi -128\right) \hat{T}^{3}+\left( -16\pi +48\right) \hat{T}%
^{2}+2\pi \hat{T} & 0\leq \hat{T}<\frac{1}{4} \\ 
\left( 128-32\pi \right) \hat{T}^{3}+\left( -144+32\pi \right) \hat{T}%
^{2}+\left( 48-10\pi \right) \hat{T}-4+\pi & \frac{1}{4}\leq \hat{T}<\frac{1%
}{2} \\ 
\left( 128-32\pi \right) \hat{T}^{3}+\left( -240+64\pi \right) \hat{T}%
^{2}+\left( 144-42\pi \right) \hat{T}-28+9\pi & \frac{1}{2}\leq \hat{T}<%
\frac{3}{4} \\ 
\left( 32\pi -128\right) \hat{T}^{3}+\left( 336-80\pi \right) \hat{T}%
^{2}+\left( -288+66\pi \right) \hat{T}+80-18\pi & \frac{3}{4}\leq \hat{T}%
\leq 1%
\end{array}%
\right.  \label{C2}
\end{equation}%
see Figs. 1 ,2, 3.

\begin{figure}[h]
\begin{equation*}
\includegraphics[width= 8 cm]{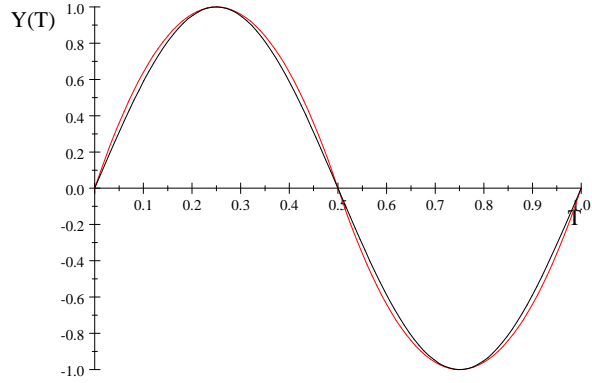}
\end{equation*}%
\caption{Displacement of the table, $Y_{q}(T)$ (black) and $Y_{s}(T)$ (red).}
\label{F1}
\end{figure}

The function $Y_{q}$, consisting of two parabolas, and its first derivative
are continuous, however its second derivative is discontinuous at $T=0,\ 
\frac{1}{2},\ 1$. The function $Y_{c_{1}}$ is smooth but provides a poorer
approximation to $Y_{s}\left( T\right) =\sin \left( 2\pi T\right) $ then $%
Y_{q}$. We have included this function because it is smooth and is the
lowest-order polynomial approximating $Y_{s}$ on the unit interval $\left[
0,\ 1\right] $.

\begin{figure}[h]
\begin{equation*}
\includegraphics[width= 8 cm]{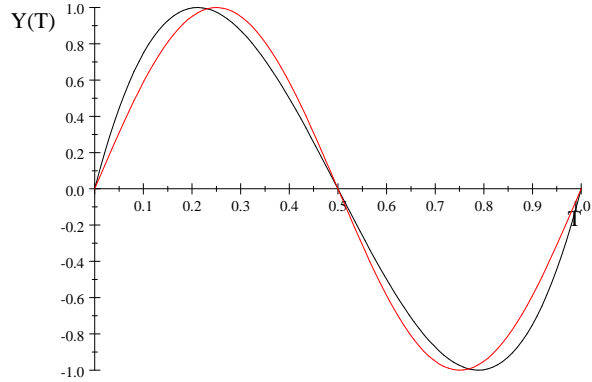}
\end{equation*}%
\caption{Displacement of the table, $Y_{c_{1}}(T)$ (black) and $Y_{s}(T)$
(red).}
\label{F2}
\end{figure}

The smooth function $Y_{c_{2}}$ consists of four cubic functions and
provides the best approximation to $Y_{s}$. Let us note that for all these
functions, $Y_{q}$, $Y_{c_{1}}$, $Y_{c_{2}}$, equation for $T_{i+1}$, cf.
Eq. (\ref{T}), is solvable.

\begin{figure}[h]
\begin{equation*}
\includegraphics[width= 8 cm]{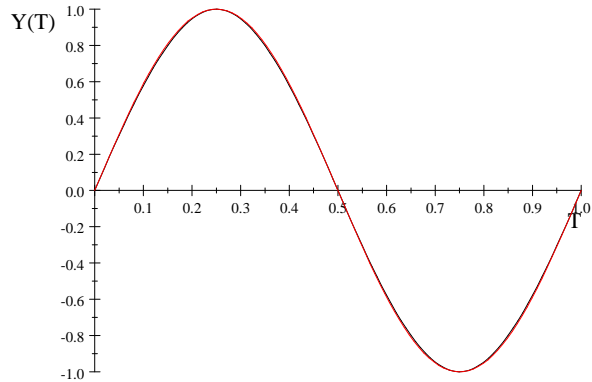}
\end{equation*}%
\caption{Displacement of the table, $Y_{c_{2}}(T)$ (black) and $Y_{s}(T)$
(red).}
\label{F3}
\end{figure}

Our model consists thus of equation (\ref{TV}) and one of Eqs.(\ref{Q}), (%
\ref{C1}), (\ref{C2}) with control parameters $R$, $\gamma $. Since the
period of motion of the limiter is equal to one, the map (\ref{TV}) is
invariant under the translation $T_{i}\rightarrow T_{i}+1$. Accordingly, all
impact times $T_{i}$ can be reduced to the unit interval $\left[ 0,\ 1\right]
$.

\section{Comparison of bifurcation diagrams}

We have computed bifurcation diagrams to study dependence of dynamics on the
model of motion of the table. Let us recall that in the case of displacement
of the table described by piecewise linear function $Y$ defined in Eq.(\ref%
{L}) the bifurcation diagram differs significantly from that computed for
sinusoidal displacement $Y_{s}$.

\begin{figure}[h]
\begin{equation*}
\includegraphics[width= 7.5 cm]{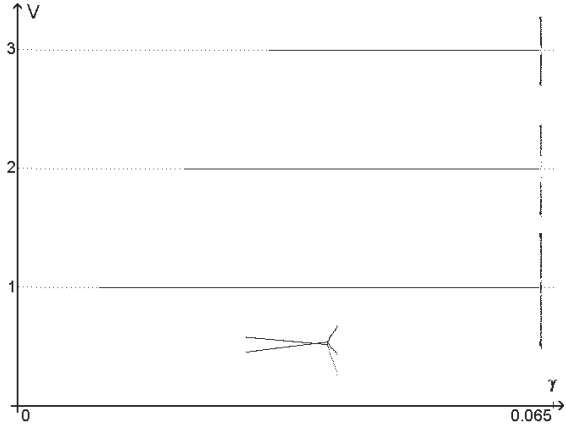}
\end{equation*}%
\caption{Bifurcation diagram, $Y(T)=Y_{q}(T)$}
\label{F4}
\end{figure}

\begin{figure}[h]
\begin{equation*}
\includegraphics[width= 7.5 cm]{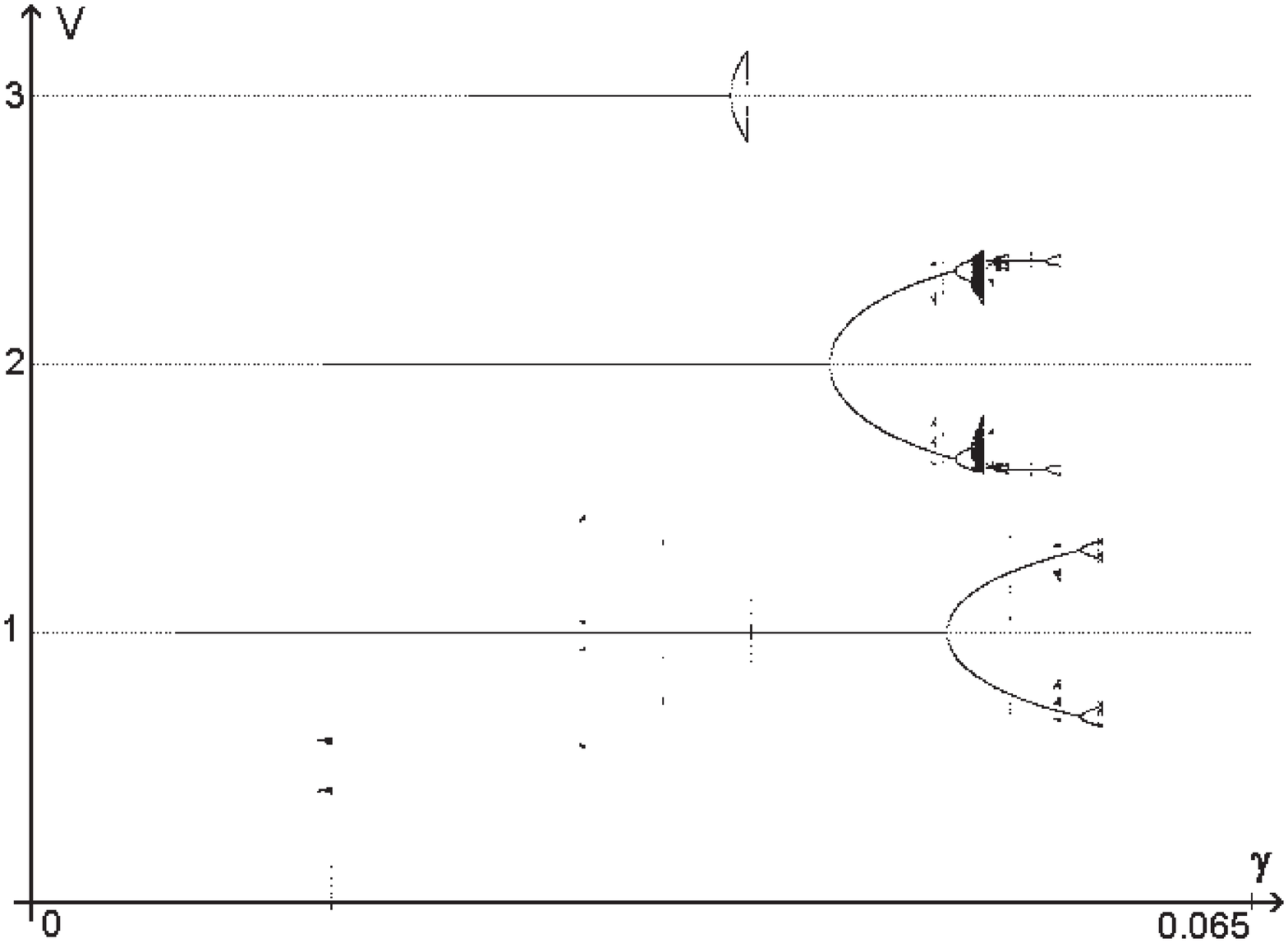}
\end{equation*}%
\caption{Bifurcation diagram, $Y(T)=Y_{c_{1}}(T)$}
\end{figure}

In the $h\rightarrow 1$ case only manifolds of periodic solutions were found
to exist and there is no chaotic dynamics \cite{AOBR2009a} while for $h\neq
1 $, classical attractors exist as well, but only one period doubling on
route to chaos via corner bifurcation was reported \cite{AOBR2009c} in
contradistinction to the case of sinusoidal displacement of the table where
full period doubling scenario is generic \cite{AOBR2007}.

The bifurcation diagram for the displacement function $Y_{q}$ is shown in
Fig. 4. There are one 2-cycle (ending after one period doubling) and six
fixed points (first three are shown) which become unstable at some $\gamma
_{cr}$ where chaotic bands appear. There are no manifolds of periodic
solutions. The bifurcation diagrams for functions $Y_{c_{1}}$ and $Y_{c_{2}}$
are presented in Figs. 5 and 6, respectively. It can be seen that full
cascades of period doubling with transition to chaos are present in Fig. 5
but it seems that some bifurcation paths end abruptly. Moreover, periodic
states loose stability in different order than for $Y_{s}$.

\begin{figure}[h]
\begin{equation*}
\includegraphics[width= 7.5 cm]{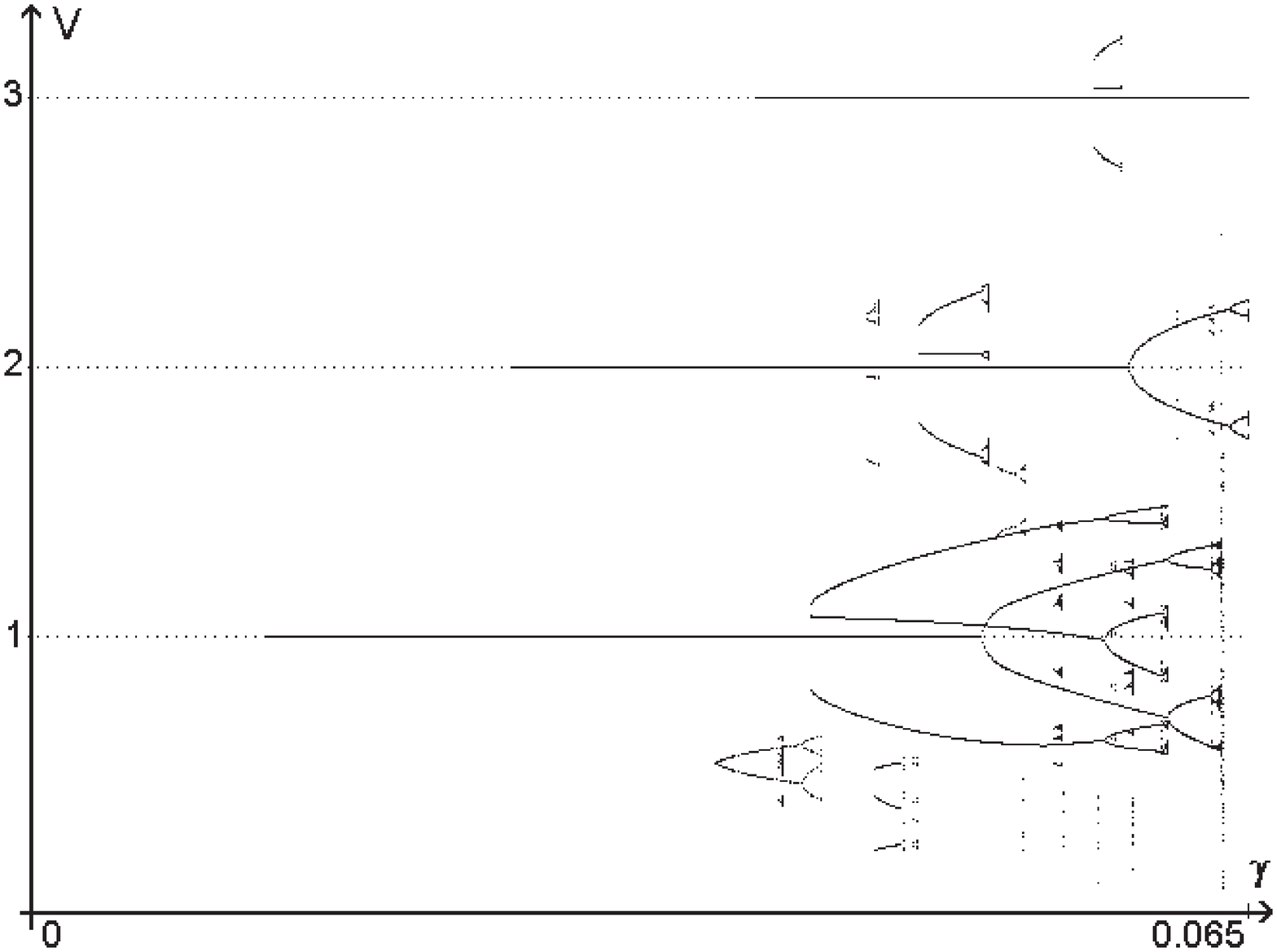}
\end{equation*}%
\caption{Bifurcation diagram, $Y(T)=Y_{c_{2}}(T)$}
\end{figure}

\begin{figure}[h]
\begin{equation*}
\includegraphics[width= 7.5 cm]{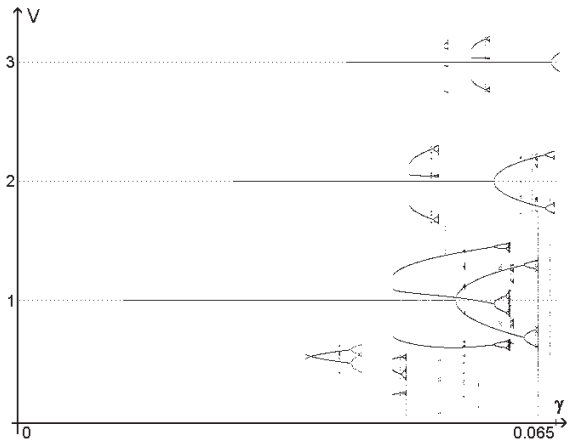}
\end{equation*}%
\caption{Bifurcation diagram, $Y(T)=sin\left( 2\protect\pi T\right) $}
\label{F7}
\end{figure}

We note finally that the bifurcation diagram computed for the displacement
function $Y_{c_{2}}$, cf. Fig. 6, is very similar to the bifurcation diagram
for the sine function $Y_{s}$, see Fig. 7.

\section{Summary and discussion}

We have constructed several simple models of table motion in bouncing ball
dynamics in order to approximate sinusoidal motion as exactly as possible.
In conclusion we can state that dynamics of the model based on Eqs. (\ref{TV}%
), (\ref{C2}) corresponds well to dynamics with table displacement given by $%
Y_{s}\left( T\right) =\sin \left( 2\pi T\right) $. Moreover, equation for
time of the next impact (\ref{T}) is a third-order algebraic equation in $%
T_{i+1}$ for $Y\left( T\right) =Y_{c_{2}}\left( T\right) $ and thus
analytical computations are possible. We are going to investigate all these
models in our future work.


\begin{thebibliography}{99}
\bibitem{diBernardo2008} M. di Bernardo, C.J. Budd, A.R. Champneys, P.
Kowalczyk, \textit{Piecewise-Smooth Dynamical Systems. Theory and
Applications}. Series: Applied Mathematical Sciences, vol. 163. Springer,
Berlin (2008).

\bibitem{Luo2006} A.C.J.Luo, \textit{Singularity and Dynamics on
Discontinuous Vector Fields}. Monograph Series on Nonlinear Science and
Complexity, vol. 3. Elsevier, Amsterdam (2006).

\bibitem{Awrejcewicz2003} J. Awrejcewicz, C.-H. Lamarque, \textit{%
Bifurcation and Chaos in Nonsmooth Mechanical Systems}.World Scientific
Series on Nonlinear Science: Series A, vol. 45. World Scientific Publishing,
Singapore (2003).

\bibitem{Filippov1988} A.F. Filippov, \textit{Differential Equations with
Discontinuous Right-Hand Sides}. Kluwer Academic, Dordrecht (1988).

\bibitem{Stronge2000} W.J. Stronge, \textit{Impact mechanics}. Cambridge
University Press, Cambridge (2000).

\bibitem{Mehta1994} A. Mehta (ed.), \textit{Granular Matter: An
Interdisciplinary Approach}. Springer, Berlin (1994).

\bibitem{Knudsen1992} C. Knudsen, R. Feldberg, H. True, Bifurcations and
chaos in a model of a rolling wheel-set. Philos. Trans. R. Soc. Lond. A 
\textbf{338} (1992) 455--469.

\bibitem{Wiercigroch2008} M. Wiercigroch, A.M. Krivtsov, J. Wojewoda,
Vibrational energy transfer via modulated impacts for percussive drilling.
Journal of Theoretical and Applied Mechanics \textbf{46} (2008) 715-726.

\bibitem{Luo2009} A. C. J. Luo, Y. Guo, Motion Switching and Chaos of a
Particle in a Generalized Fermi-Acceleration Oscillator, Mathematical
Problems in Engineering, vol. \textbf{2009}, Article ID 298906, 40 pages,
2009. doi:10.1155/2009/298906.

\bibitem{AOBR2008} A. Okninski, B. Radziszewski, Dynamics of impacts with a
table moving with piecewise constant velocity, Vibrations in Physical
Systems, vol. XXIII, p.289 -- 294, C. Cempel, M.W. Dobry (Editors), Pozna%
\'{n} 2008.

\bibitem{AOBR2009a} A. Okninski, B. Radziszewski, Dynamics of a material
point colliding with a limiter moving with piecewise constant velocity, in:
Modelling, Simulation and Control of Nonlinear Engineering Dynamical
Systems. State-of-the Art, Perspectives and Applications, J. Awrejcewicz
(Ed.), Springer 2009, pp. 117-127.

\bibitem{AOBR2009b} A. Okninski, B. Radziszewski, Dynamics of impacts with a
table moving with piecewise constant velocity, Nonlinear Dynamics \textbf{58}
(2009) 515-523.

\bibitem{AOBR2009c} A. Okninski, B. Radziszewski, Chaotic dynamics in a
simple bouncing ball model, Proceedings of the 10th Conference on Dynamical
Systems: Theory and Applications, December 7-10, 2009. {\L }\'{o}d\'{z},
Poland, J. Awrejcewicz, M. Kazmierczak, P. Olejnik, J. Mrozowski (eds.), pp.
651-656.

\bibitem{AOBR2010} Chaotic dynamics in a simple bouncing ball model,
arXiv:1002.2448 [nlin.CD].

\bibitem{AOBR2007} A. Okninski, B. Radziszewski, Grazing dynamics and
dependence on initial conditions in certain systems with impacts,
arXiv:0706.0257 (2007).
\end{thebibliography}
\end{document}